\documentclass[fleqn,10pt]{wlscirep}
\usepackage[utf8]{inputenc}
\usepackage[T1]{fontenc}

\usepackage{cite}
\usepackage{amsmath,amssymb,amsfonts,amsthm}
\usepackage{algorithmic}
\usepackage{graphicx}
\usepackage{textcomp}
\usepackage{xcolor}
\usepackage{siunitx}
\usepackage{comment}
\usepackage{caption}
\usepackage{subcaption}
\usepackage{tabularx}
\usepackage{multirow}

\newtheorem{example}{Example}
    
\begin{document}

\title{Modular and Extendable\\1D-Simulation for Microfluidic Devices}

\author{Maria Emmerich,$^{1,*}$ Florina Costamoling,$^{2}$ and Robert Wille$^{1,3,*}$ \vspace{.3cm}\\ $^{1}$~Technical University of Munich (TUM), Arcisstrasse 21, 80333 Munich, Germany \\$^{2}$~Johannes Kepler University Linz (JKU), Altenberger Strasse 69, 4040 Linz, Austria \\$^{3}$~Software Competence Center Hagenberg GmbH (SCCH), Softwarepark 32a, 4232 Hagenberg, Austria \\ $^{*}$ maria.emmerich@tum.de, robert.wille@tum.de \\ https://www.cda.cit.tum.de/research/microfluidics/}


\begin{abstract}
Microfluidic devices have been the subject of considerable attention in recent years. The development of novel microfluidic devices, their evaluation, and their validation requires simulations. 
While common methods based on \emph{Computational Fluid Dynamics}~(CFD) can be time-consuming, 1D simulation provides an appealing alternative that leads to efficient results with reasonable quality.
Current 1D simulation tools cover specific microfluidic applications; however, these tools are still rare and not widely adopted. There is a need for a more versatile and adaptable tool that covers novel applications, like mixing and the addition of membranes, and allows easy extension, resulting in one comprehensive 1D simulation tool for microfluidic devices.
In this work, we present an open-source, modular, and extendable 1D simulation approach for microfluidic devices, which is available as an open-source software package at https://github.com/cda-tum/mmft-modular-1D-simulator.
To this end, we propose an implementation that consists of a base module (providing the  core functionality) that can be extended with dedicated \mbox{application-specific} modules (providing dedicated support for common microfluidic applications such as mixing, droplets, membranes, etc.). Case studies show that this indeed allows to \emph{efficiently} simulate a broad spectrum of microfluidic applications in a quality that matches previous results or even fabricated devices.
\end{abstract}

\flushbottom
\maketitle

\thispagestyle{empty}

\section{Introduction}\label{sec:Intro}
The use of microfluidic devices has made substantial contributions across diverse industries, including the development of medical testing devices, the development and analysis of pharmaceuticals, chemical synthesis, and many more \cite{nguyen_1_2019}. Microfluidic devices for each of these applications are tailored to their specific use cases, leading to a variety of different designs. With this increase in capabilities, the complexity of properly and correctly designing them increases as well~\cite{bennett_microfluidic_2009}.

In fact, small changes, e.g., in the dimensions of channels, the pressure of the pumps, or the flow rate of the fluid can significantly impact the behavior of the whole device~\cite{takken_simulation_2022}. Still, the microfluidics industry follows an iterative \mbox{trial-and-error} approach that relies on personal expertise, manual calculations, and the manufacture of multiple prototypes~\cite{fink_accurate_2021}---eventually resulting in an \mbox{error-prone}, costly, and \mbox{time-consuming} design process~\cite{liu_design_2022, huang2021computer}.

Simulation methods can help to overcome this complexity and speed up the design process. They allow to test and review the validity of a given design, to design parts of a device without a fabricated prototype~\cite{carvalho_computational_2021}, or to evaluate different possible implementations to determine a design which is most robust. But obviously, the simulation of complex microfluidic behavior comes at a cost which heavily depends on the degree of abstraction from the real-world behavior. 

In fact, the following abstraction can be applied: 
\begin{itemize}
    \item \emph{Actual Physical Device}, which obviously constitutes the most accurate ``representation''; but prototype fabrication is error-prone, costly, and time consuming.
    \item \emph{Computational Fluid Dynamics}~(CFD), a simulation method that provides high accuracy, but is computationally expensive \cite{oh2012design}.
    \item \emph{Compartment Models}, a high abstraction model,
    which separates the system into functional zones called compartments \cite{jourdan_compartmental_2019} and simulates each compartment 
    independently (connected by predefined fluxes), but the set-up requires high effort.
    \item \emph{1D Simulation}, a different high-abstraction model that simplifies the system into an analytical solution \cite{takken_simulation_2022}.
\end{itemize}

While CFD is often used, it leads to long simulation times due to the computational load, and higher abstractions need to be employed \cite{takken_simulation_2022}. In compartment models, the compartments are simulated separately, resulting in the consideration of the behavior of the whole system during compartmentalization, but it is eventually neglected in the simulation. In complex microfluidic networks, this can lead to a build-up of errors. 

Instead, 1D simulation allows to handle much more complex systems efficiently while, at the same time, providing results with meaningful quality---making them an appealing alternative. 
1D abstraction models are available for multiple microfluidic components and processes, but simulation tools are still rare and not widely adopted. The use of electric circuit simulation tools to simulate microfluidic devices is not practical, and the currently available microfluidic tools only support very specific microfluidic applications. In fact, essential applications such as mixing or the inclusion of membranes are not supported by any 1D simulation approach so far. 

In this work, we are addressing this problem by proposing an open-source, modular, and extendable 1D simulation approach for microfluidic devices. The main idea is to employ a base module as the foundation, which provides the core functionality for 1D simulation. On top of that, several \mbox{application-specific} modules can be added that provide dedicated support. This results in a simulation approach that utilizes the 1D abstractions (hence, providing an efficient simulation) but remains applicable to several microfluidic applications, some of which can be supported for the first time.

The advantages of the implemented module extensions are confirmed by case studies 
covering several real-world examples from published works and showcasing that the proposed approach indeed allows to \emph{efficiently} simulate a broad spectrum of microfluidic applications in a quality that matches previous results or even fabricated devices and thereby validates the simulations.

The remainder of this paper is structured as follows: First, we review 1D simulation and its current limitations in detail--providing the motivation for this work. Then, in Section~\ref{sec:Components} we introduce the modular and extendable approach proposed in this work---covering its base module and three representative application-specific modules. These are then evaluated in Section~\ref{sec:cases} by using "real-world" examples from published works. Finally, Section \ref{sec:concl} concludes this paper.

\section{Motivation}
The design and validity of microfluidic devices can be verified by using simulation methods. To optimize simulation time, 1D~abstraction can be employed as an alternative to classical CFD simulations. In this section, we review the core concepts of 1D~simulation and discuss its limits, which motivate this work.

\subsection{1D Simulation}\label{sec:1dSim}
The 1D model can be applied in settings with a laminar flow regime, a fully developed, viscous, and incompressible fluid flow. Subsequently, this abstraction can be efficiently applied to microfluidic networks and leads to a \mbox{hydraulic-electric} circuit analogy, where microfluidic networks can be simulated similarly to electric circuits. Compared to CFD simulations, this allows for a faster and easier setup~\cite{fink_accurate_2021}, faster refinement~\cite{voigt_method_2020}, and significantly reduced simulation runtimes for complex networks~\cite{voigt_method_2020, takken_simulation_2022}.

More precisely, in flow-based systems, the dependency of pressure drop~$\Delta P$, volumetric flow rate~$Q$, and hydrodynamic resistance~$R$ can be described by \mbox{\emph{Hagen-Poiseuille's law}} \mbox{$\Delta P = Q \times R$}~\cite{oh2012design, bruus2008theoretical}, which is analogous to \mbox{\emph{Ohm's law}} in the electrical domain. This allows for the translation from a microfluidic network to an electric circuit. 
Accordingly, the volumetric flow rate in the microfluidic network corresponds to the flow of electricity (electric current) and the channel resistance for rectangular channels, with $h/w < 1$, described~by 
\begin{equation}\label{eq:resistance}
    R(l)=12 \left[ 1- \frac{192 h}{\pi ^5 w} tanh \left( \frac{\pi w}{2h} \right) \right] ^{-1} \frac{\mu _c l}{w h^3},
\end{equation}
corresponds to the electric resistance~\cite{glawdel2012global}. 
Since most channel geometries in microfluidic networks are rectangular~\cite{oh2012design}, this expression is widely applicable, while hydraulic resistances for other channel geometries are described elsewhere~\cite{bruus2008theoretical, mortensen_reexamination_2005}. These are not included in the current implementation but can be adapted in the underlying code by interested researchers, although such modifications would require further testing and validation to ensure accuracy. 
Flow or pressure pumps are equivalent to independent, constant current or voltage sources, respectively, and mass conservation correlates with the current law and energy conservation with the voltage law.

To include time-dependent events in the microfluidic network, the system can be continuously updated. This way, changes in concentration, or flow rate, or movement of, e.g., droplets, can be included in the simulation.

% \begin{figure} [t]
% 	\centering
% 	 	\begin{subfigure}[b]{0.25\linewidth}
% 		\centering
% 		\includegraphics[width=\textwidth]{Images/ElectricCircuitAnalogyA_New.png}
% 		\caption{Microfluidic Network}
% 		\label{fig:MicrofluidicNetwork}
% 	\end{subfigure}
%   	\begin{subfigure}[b]{0.25\linewidth}
% 		\centering
% 		\includegraphics[width=\textwidth]{Images/ElectricCircuitAnalogyB_New.png}
% 		\caption{Electric Circuit}
% 		\label{fig:ElectricCircuit}
% 	\end{subfigure}
% 	\caption{Electric circuit analogy}
% 	\label{fig:MF_network_electric_circuit}
% \end{figure}

\begin{figure} [t]
    \centering
    \includegraphics[width=0.7\linewidth]{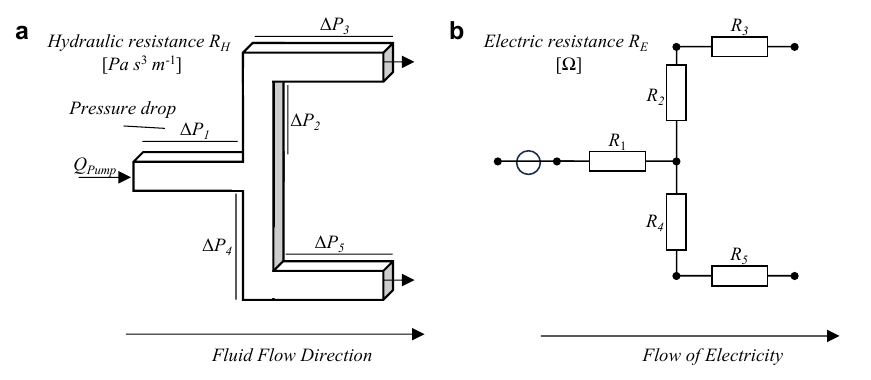}
    \caption{Electric circuit analogy, \textbf{a} Microfluidic network, \textbf{b} Electric circuit}
    \label{fig:MF_network_electric_circuit}
\end{figure}

\begin{example}
    Consider the microfluidic network shown in Figure \ref{fig:MF_network_electric_circuit}a. This network can be translated into an electric circuit, as shown in Figure \ref{fig:MF_network_electric_circuit}b. Each channel in the microfluidic network corresponds to a resistance in the electric circuit, whereas the channel branching is represented by nodes. The flow rate pump of the microfluidic network corresponds to an independent, constant current source in the electric circuit.
\end{example}

This geometric abstraction of 1D simulation has immense potential for microfluidics. However, it has yet to be fully utilized. The current limits of the 1D simulation are described in the next section.

\subsection{Limits of Current 1D Approaches}
Despite the advantages of 1D simulation, current 1D simulation tools have several limitations that impede their usability and versatility. 

Simple microfluidic networks have been manually translated to a 1D electric circuit representation and specified using classical electronic design automation tools \cite{voigt_method_2020, wang_microfluidic_2010, oh2012design, zaidon_modelling_2015, han_computer-aided_2019}. However, the use of electronic design automation tools requires high effort for the manual translation of microfluidic components to electric circuits and potential additional specifications of component behavior by the user. This translation step needs to be redone for each microfluidic design refinement. And while it can be useful for experts that are already familiar with their usage, experts from the microfluidic domain and its adjacent fields are most likely not familiar with their usage, highlighting the need for dedicated microfluidic simulation tools.

In some cases, dedicated 1D simulation tools for specific microfluidic applications have been developed, e.g., the \emph{Munich Microfluidics Toolkit} (MMFT) droplet simulator \cite{fink_mmft_2022}, which is able to simulate droplet microfluidics on a 1D abstraction level. An alternative is the SS-Analyzer that can convert and simulate designs created with the continuous flow microfluidic design tool 3d$\mu$F \cite{sanka20193d}, but is limited to \mbox{1-1} connections and mixer objects created in 3d$\mu$F.
However, while these tools show the potential of 1D simulation, they are only functional for the specific microfluidic applications for which they were developed.

Therefore, any other or novel microfluidic application currently requires the development of its own unique 1D simulation tool from scratch.
This constant redevelopment of a new simulators for each use case is redundant and infeasible.
Hence, there is a need for a comprehensive 1D simulation method that is capable of simulating various complex microfluidic networks, including different applications such as continuous flow, droplets, mixing, or the addition of membranes.

In this work, we propose such a modular and extendable 1D simulation approach. This resulting simulator is able to represent microfluidic devices at a high abstraction level, allowing for an efficient simulation, and, at the same time, provide further modules that can properly cover additional applications such as mixing, droplets, and membranes---substantially broadening the application of existing 1D simulation tools. 

% \begin{figure*}[t]
% 	\centering
%  	\begin{subfigure}[b]{0.43\textwidth}
% 		\centering
% 		\includegraphics[width=\textwidth]{Images/BaseModule.png}
% 		\caption{Base}
% 		\label{fig:BaseModule}
% 	\end{subfigure}
%   	\begin{subfigure}[b]{0.43\textwidth}
% 		\centering
% 		\includegraphics[width=\textwidth]{Images/MixingModuleWhite.png}
% 		\caption{Mixing}
% 		\label{fig:MixingModule}
% 	\end{subfigure}
%    	\begin{subfigure}[b]{0.43\textwidth}
% 		\centering
% 		\includegraphics[width=\textwidth]{Images/DropletModule.png}
% 		\caption{Droplets}
% 		\label{fig:DropletModule}
% 	\end{subfigure}
% 	\begin{subfigure}[b]{0.43\textwidth}
% 		\centering
% 		\includegraphics[width=\textwidth]{Images/MembraneModule_New.png}
% 		\caption{Membranes}
% 		\label{fig:MembraneModule}
% 	\end{subfigure}
%         \caption{Capabilities of the 1D Simulation}
%         \label{fig:Modules}
% 	\hfill
% \end{figure*}

\begin{figure} [t]
    \centering
    \includegraphics[width=0.85\linewidth]{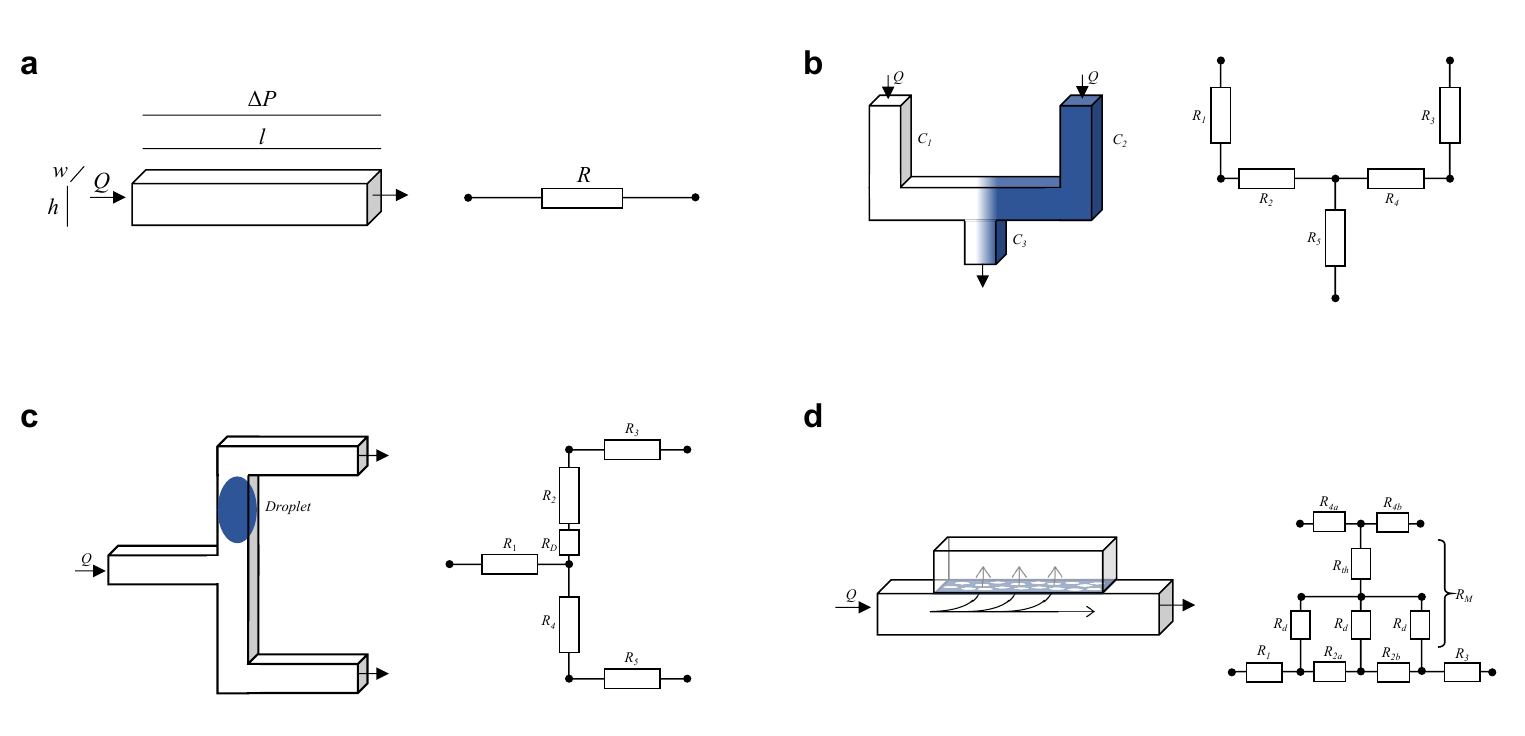}
    \caption{Capabilities of the 1D Simulation, \textbf{a} Base, \textbf{b} Mixing, \textbf{c} Droplets, \textbf{d} Membranes}
    \label{fig:Modules}
\end{figure}

\section{Proposed Modular~and~Extendable~1D~Simulation}\label{sec:Components}
In this section, we describe the proposed solution in detail. To this end, we examine three key microfluidic applications---mixing, droplets, and membranes---as representative examples to showcase the flexibility of the approach. 
This way, we consider applications that have been considered separately before (namely droplets), as well as applications which, for the first time have been represented in 1D (namely mixing and membranes)---providing a wide range of module extensions for this simulation approach.
Figure \ref{fig:Modules} illustrates the corresponding use cases as well as their 1D formulation (represented through the electric circuit analogy). In the following, each application is
described in detail, starting with the base module.

\subsection{Base Module}
The base module serves as the \emph{simulation engine} and defines the network structures upon which various modules can be build. The simulation engine manages the microfluidic network representation and advances the simulation through computations and events. Computations update the network by recalculating the pressure at each node using \emph{Modified Nodal Analysis}~(MNA)~\cite{najm_circuit_2010}, while events represent incidents that require special treatment and determine the time steps at which the simulation state is recalculated.

\begin{figure} [t]
	\centering
	\includegraphics[width=0.7\linewidth]{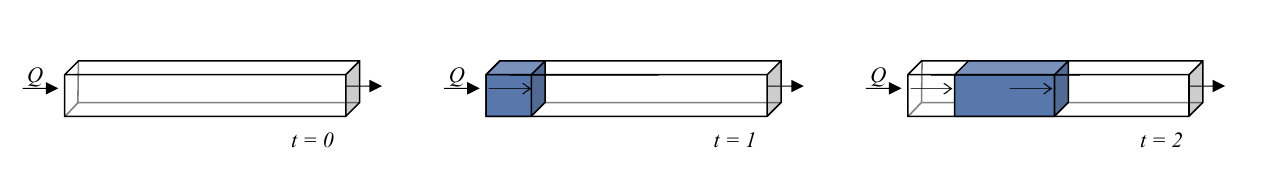}
	\caption{Mixture or Fluid Movement based on Time Steps}
	\label{fig:TimeSteps}
\end{figure}

The foundation of the base module of channel-based microfluidic devices is the simulation of continuous fluid flow. 
This simple flow can be easily described in the 1D model, as reviewed in Section~\ref{sec:1dSim}.
To capture dynamic behavior, such as the movement of specific fluids or droplets in the channel network, \emph{discrete time steps} can be included. More precisely, a time step event is defined as the the minimum of either a specified input or the shortest time $t$ it takes the fluid to flow through a channel in the network, i.e., \mbox{$t=V_{Channel}/Q_{Channel}$}.
The time step event leads to an update of the simulation state based on the previously calculated values. Subsequently, variations in the inlet concentration can be simulated, as well as their movement through the system. In Figure~\ref{fig:TimeSteps}, this fluid (or mixture) movement is sketched for three time step events. The fluids move sequentially through the channel network. Due to the convection-dominated flow regime, lateral diffusion is neglected~\cite{ronaldson-bouchard_multi-organ_2022}. 
Overall, this provides the basis for all 1D simulations.

\begin{example}
    Consider the channel shown in Figure \ref{fig:Modules}a. The fluid in the channel is defined by the channel inflow and its flow rate by the hydraulic resistance. If another fluid flows into the channel, both fluids move in sequence through the channel.
\end{example}

The base module can be extended to integrate more functionality through additional modules. These extensions can either work independently of other modules--like the droplet module, or the mixing module--or by building on top of existing module functionality, such as the membrane module, which utilizes the mixing module's functionality. This modular design allows for new applications to be added on top of the base module, either using already implemented module functionalities when needed or keeping them separate. In the following sections, the three already implemented modules are described.

\subsection{Mixing}
In microfluidic devices, mixing is one of the key operations required for most assays. Two different fluids or fluids that contain different concentrations can be mixed at a defined rate to achieve the desired final concentration.

This can be integrated on top of the base module by extending the definition of fluids as mixtures, i.e., the volume inside the channel is no longer defined as a fluid containing a single species but as a mixture that can contain several fluids or species. This allows for the mixing of the liquid volumes within the channel network. The mixing and the resulting concentration~$C_0$ at the outlet are then defined by the flow rates~$Q_i$ and the concentration in the inflow channels~$C_i$, i.e., 

\begin{equation}\label{eq:permeability}
    C_0 = \frac{\sum C_i \times Q_i}{\sum Q_i},
\end{equation} where~$C_0$ can usually be determined by defining the flow rates of the inflow channels, as the concentration values of the inflow fluids are usually predefined \cite{fink2022automatic}. This model assumes non-reactive, fully miscible fluids, meaning reactions between species during the simulation are not considered and the fluids mix completely without separating into phases. Furthermore, this model assumes instantaneous mixing at the node, meaning fluids fully mix upon entering the node. Nevertheless, the respective ratios of mixed fluids are accurately accounted for at each node, ensuring reliable predictions for mixing outcomes for long enough channels. Further developments could extend this model to incorporate species reactions or diffusive mixing along the channel.

\begin{example}
    Consider the mixing operation shown in Figure~\ref{fig:Modules}b. Two channels flow into a third channel. Both contain different concentrations or fluids. When they flow into the third channel, the fluids mix. In the 1D model (represented as an electric circuit in Figure \ref{fig:Modules}b), each channel is defined through its resistance, which defines the flow rate. From that and the fluid concentrations, the resulting mixing ratio can be determined. More precisely, the white and blue fluids flow into the same channel at equal flow rates, resulting in a 50/50 mixture.
\end{example}

\subsection{Droplets}\label{sec:droplets}
In droplet-based microfluidics the manipulation of droplets is utilized for chemical and biological assays~\cite{postek_droplet_2022}.
They have been simulated using dedicated 1D simulation tools before~\cite{fink_mmft_2022}. However, these tools lack the functionality to simulate any other microfluidic applications than those for which they were specifically designed. Here, the simulation of droplet microfluidics is added as another module extension to create a more versatile simulation method. 

For this, the droplet behavior inside the device is defined and updated based on pressure drops and flow rates in the microfluidic network. Droplets can be injected at the inlets, but the model does not simulate their formation. Instead, it focuses on the movement through the microfluidic network. 
The determination of droplet movement through the network is described in more detail in~\cite{fink_mmft_2022}. In short, first, the current flow rates and, then, all potential next events are computed. The droplets are then moved based on the flow rates in the channel until the next event would be triggered. After this, the event is performed and the current flow state is updated. This is repeated until the simulation terminates because, e.g., all droplets have left the network. Alternative models define the droplet speed separately to the flow rate in the channel, meaning flow rate and droplet speed are not necessarily identical~\cite{jakiela_speed_2011, balestra_viscous_2018}. Both approaches are supported by empirical data, but we opted for the model where the droplet resistance directly impacts the total channel resistance.

More precisely, as proposed in~\cite{glawdel2012global}, when adding a droplet of length $l_{Droplet}$ to a channel, the droplet resistance is defined as  
\begin{equation}
    R_{Droplet} = b \frac{a \mu _c l_{Droplet}}{w h^3}
\end{equation} where $b$ is a factor that depends on the experimental setup and usually lies between 2 and 5. The resistance of a channel that contains $n$ droplets then has a total resistance of $R^*= R + n R_{Droplet}$.

\begin{example}
    Consider Figure \ref{fig:Modules}c. The position of the droplet based on its movement increases the resistance in a channel. This is realized by defining the droplet itself as a resistance. The number of time steps that a droplet remains in a channel is known based on the flow rate in the channel. Once it exits one channel, the network state changes. This triggers an event and a recalculation of the network state, including the effect of the new droplet position on the resistances and flow rates. Overall, this allows for the determination and evaluation of the path that all droplets take through a network, including their position during the course of the simulation time.
\end{example}

\subsection{Membranes}
The functionality of microfluidic devices can be increased by adding complex geometries, not only restricted to channels. Species transport (including the species concentrations contained in a fluid) can be selectively altered by using porous or semi-permeable membranes.

More precisely, the species diffusion is defined by Fick's first law, i.e., 
\begin{equation}\label{eq:fickslaw}
    J_i=-D_i\frac{\delta C_i}{\delta x},
\end{equation}
based on the species flux~$J_i$, the diffusion factor of the species $D_i$, the concentration~$C_i$, and the distance~$x$.
Including the mass balance, the time-dependent concentration change in the tank $\frac{\delta C_{i,t}}{\delta t}$ is dependent on the tank's volume~$V_t$, the membrane surface $A_m$, the species-dependent membrane permeability~$P_{i, m}$, as well as the concentration difference between channel and tank~$\Delta C_i$~\cite{ronaldson-bouchard_multi-organ_2022}, i.e.,
\begin{equation}\label{eq:massbalance}
    V_{t}\frac{\delta C_{i,t}}{\delta t}=A_{m} \times P_{i, m} \times \Delta C_i.
\end{equation}

The permeability $P_{i, m}=1/R_{M}$ can be calculated by abstracting the membrane geometry to a resistance value. Multiple resistance models exist and are implemented in the simulator~\cite{berg_random_1993, chung_use_2018, vandersarl_rapid_2011, ronaldson-bouchard_multi-organ_2022}. Choosing the right model is crucial to obtain the correct results. In the following, we focus on the model based on the pore discovery $R_d=1/(4rD_F)$, the number of pores $N_p=A_m/A_p\times porosity$ \cite{berg_random_1993, chung_use_2018}, and the resistance of the effective thickness the species need to cross $R_{th}=th/(A_m \times D_F)$ adapted from \cite{berg_random_1993}, i.e., 
\begin{equation}\label{eq:PoreR}
    R_M=\frac{R_d}{N_p} + R_{th}.
\end{equation}
Since this resistance model already includes the area and the permeability, the flux equation translates to \mbox{$J_{tank-channel}=1/R_{M}\Delta C_i$}.

The partial differential equation from Eq.~\ref{eq:fickslaw} can be solved using the well-known \emph{classic Runge-Kutta method} (RK4). Conveniently, the simulator already updates the solution at regular time steps. The flux is calculated at each time step based on the previous concentration difference $\Delta C$ and added to, or subtracted from, the concentration in the tank and channel, respectively. 

\begin{example}
Consider the membrane shown in Figure \ref{fig:Modules}d. It is attached to the side of the channel. On the other side of the membrane, a tank-like geometry is defined. The fluid flow into this tank is restricted by the membrane. Only species that can permeate the membrane are transported across. These species flow in and out of the tank again according to the membrane resistance and concentration.
\end{example}

\begin{figure}[t]
    \centering
    \includegraphics[width=0.4\linewidth]{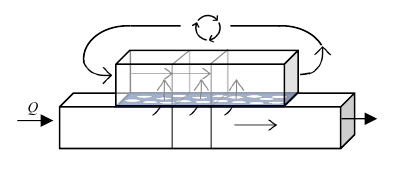}
    \caption{Mixture Movement}
    \label{fig:MixtureMovement}
    \hfill
\end{figure}

In Figure \ref{fig:MixtureMovement}, the distribution and movement of the species is depicted.
To prevent the forward propagation of species in the tank, when assuming an ideally mixed compartment the mixture separation of the channel is mirrored in the tank. The three mixtures in the channel in Figure \ref{fig:MixtureMovement}, have the same position in the tank above the channel. There is no advective flow in the compartment, but to better model the species distribution, the mixtures in the tank are moved alongside the channel mixtures. When they reach the end of the tank, instead of being discarded, the mixtures are "recycled" and added to the front of the tank, symbolized by the arrows above the tank in Figure \ref{fig:MixtureMovement}. 
This way, the law of mass conservation is obeyed. 

\begin{example}
Consider again Figure \ref{fig:MixtureMovement}, in an ideally mixed tank, only one fully mixed mixture would be present. When there is more than one mixture below the tank, species that are diffusing into the tank through the membrane from the last mixture in the channel (on the left) could then diffuse back out of the tank into the first mixture (on the right), effectively jumping ahead in the channel. To address this, the tank contains the mirrored mixtures of the channel. Instead of flowing out of the tank, the last mixture is added to the front of the tank. This ensures that no species are deleted in the course of the simulation.
\end{example}

\section{Application and Case Studies}\label{sec:cases}

The modular and extendable simulation approach as proposed above has been implemented in C++ and is now available as open-source tool at https://github.com/cda-tum/mmft-modular-1D-simulator, along with test files, including additional network definitions, and a step-by-step guide. Several predefined simulation cases, including the following case studies, are provided for each module as test files and can serve as baselines for users to define their individual designs. This allows to focus on adapting relevant parameters, without the need for setting up new simulation cases. For the first time, this allows for efficient 1D simulations of various microfluidic applications (some of which have not been supported in 1D yet) within one tool. If required, more modules can be added. To demonstrate the applicability as well as the accuracy of the resulting tool, in the following three case studies are presented. We have compared our results to either published experimental data or CFD simulations or both, depending on the case study as one of the most important aspects of evaluating simulation approaches is the verification of accuracy. 
More precisely, we consider a gradient generator~\cite{fink2022automatic}, i.e., a mixing method, a droplet ring network~\cite{fink2020automatic}, and membrane-based experiments~\cite{chramiec_integrated_2020}, 
as representative examples of microfluidic devices.
In each case, we can show that the proposed 1D simulation method is readily available and can conduct this broad range of simulations in negligible runtime while, at the same time, producing results that match the quality of previous simulated results or even fabricated devices for the mixing and membrane case studies. 

\begin{figure}[t]
	\centering
	\includegraphics[width=0.6\linewidth]{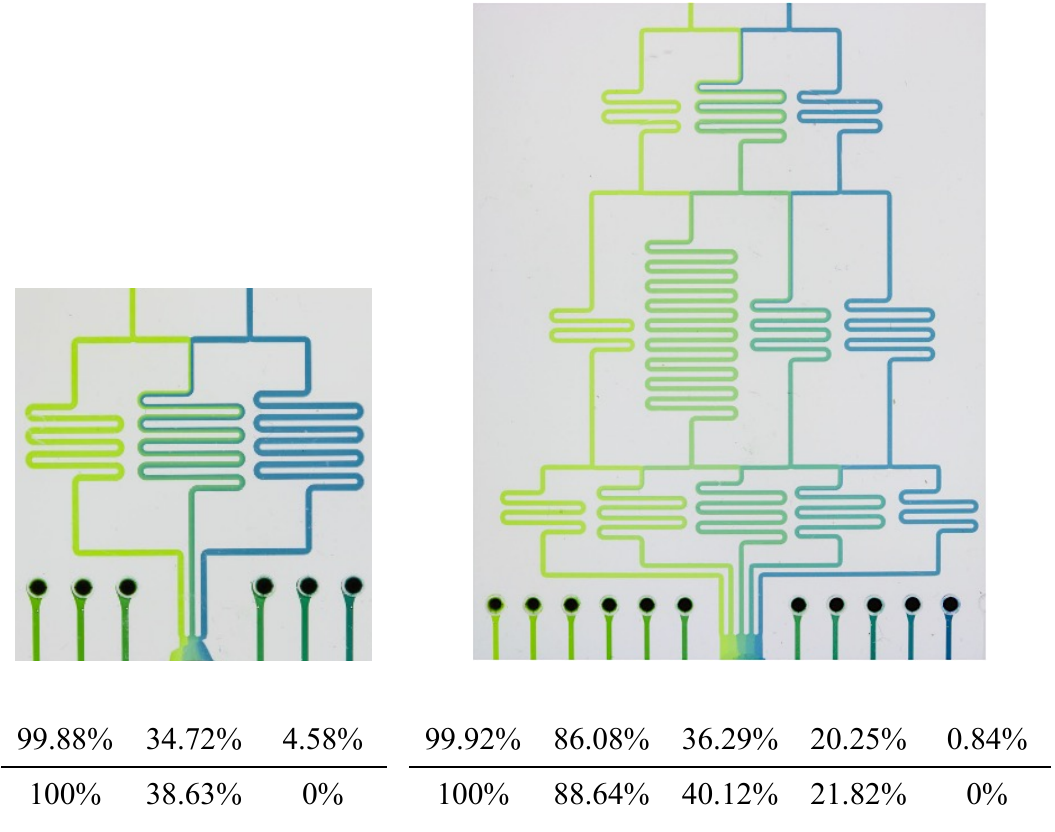}
	\caption{Gradient Generators~\cite{fink2022automatic} and the concentration at the outlets (top row: experimentally measured, taken from~\cite{fink2022automatic}; bottom row: simulator results)}
	\label{fig:CaseStudyMixing}
\end{figure}

\subsection{Mixing}
As a first case study, we consider a gradient generator. Those are microfluidic devices which conduct several mixing steps in order to create desired concentrations. To this end, inlet concentrations, channel geometries, as well as the applied flow rates are utilized \cite{fink2022automatic}. The top of Figure~\ref{fig:CaseStudyMixing} shows two gradient generators which have been considered: One with three outlets that are supposed to generate outlet concentrations of 100~\%, 38.63~\%, and 0~\%; and another with five outlets that are supposed to generate outlet concentrations of 100~\%, 88.64~\%, 40.12~\%, 21.82~\%, and 0~\%. Specifications for both devices have been taken from~\cite{fink2022automatic}.

Based on these specifications, we generated the corresponding designs and used them as inputs to the proposed simulation approach. Afterwards, we compared the resulting concentrations at the outlets (obtained by the simulator) with those measured at the actually fabricated device (as reported in~\cite{fink2022automatic}). The results are shown in the bottom table of Figure~\ref{fig:CaseStudyMixing}. They confirm the quality that, despite the abstraction, can be obtained by the proposed simulation approach. In fact, the results the simulator generates for the gradient generator with three outlets, the maximal absolute deviation is 4.58~\% and the smallest 0.12~\%. For the gradient generator with five outlets, the maximal absolute deviation is 3.83~\% and the smallest 0.08~\% compared to the values obtained from the fabricated device, resulting in a mean absolute error~(MAE) of 2.19~\%. 
The deviation between the fabricated device and the simulation can be, as stated in the original paper, attributed to tolerances in the fabrication process \cite{fink2022automatic}. 

\begin{figure}[t]
	\centering
	\includegraphics[width=0.5\linewidth]{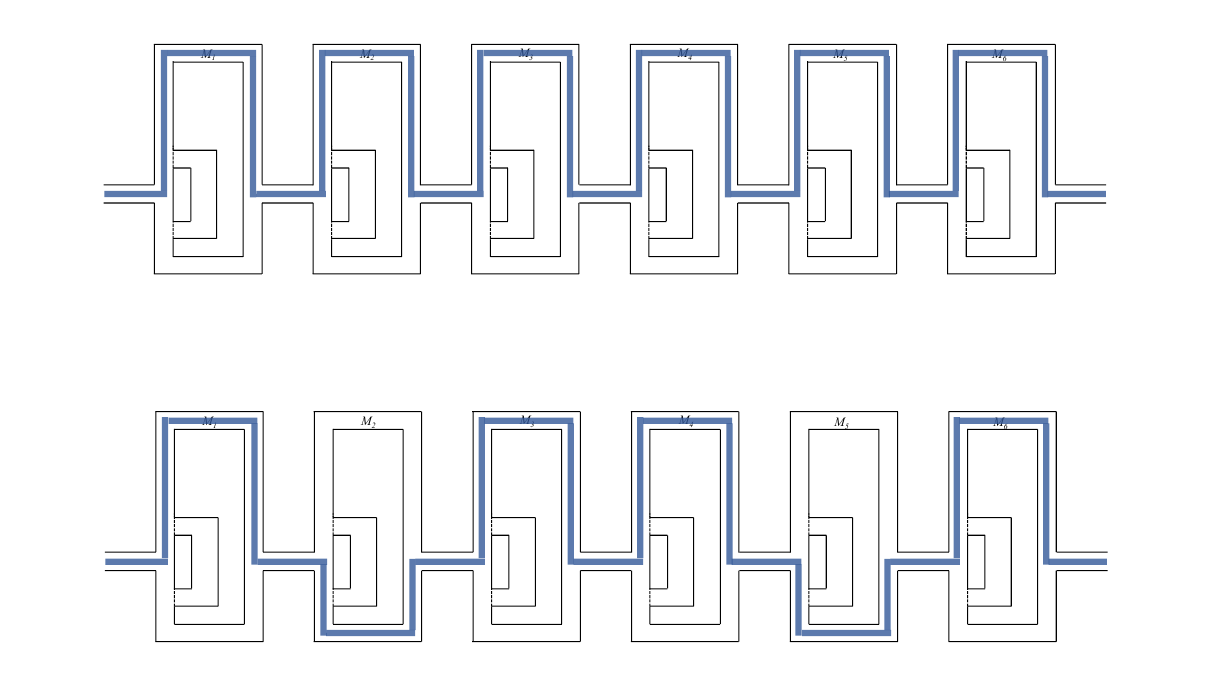}
    \vspace{0.2cm}
	\caption{Simulated payload droplet routing paths based on example networks that were verified using CFD simulations from~\cite{fink2020automatic}}
	\label{fig:CaseStudyDroplets}
\end{figure}

\subsection{Droplets}
In a second case study, we considered time-sensitive ring network designs as proposed in \cite{fink2020automatic}. 
In ring networks, it is possible to address one or more modules in a network by using a \mbox{so-called} payload droplet. This could, e.g., allow a biological sample to be routed through designated unit operations for DNA sequencing or cell analysis \cite{teh2008droplet}. To route this payload droplet as desired, hydrodynamic effects are utilized. Droplets choose the path of least resistance (or highest flow rate)  at an intersection and lead to an increase of channel resistance (and drop in flow rate) in the channel they are located in. This way, droplets impact each other's paths. By using \mbox{so-called} header droplets (that contain no sample and are only used to route the payloads), it is possible to selectively adapt the channel resistances and direct the payload droplet through the network as desired~\cite{fink2020automatic}.

For the case study, we considered the design from~\cite{fink2020automatic}, including two exemplary droplet sequences.
For the simulation, the distances were translated into a timed sequence of injections. The injection of the first droplet occurs at the start of the simulation~$t_0 = 0s$. The injection times~$t_i$ for all subsequent droplets are calculated based on the given droplet distance~$d_i$, the volumetric flow rate of the pump~$Q_{in}$, and the width~$w_{c1}$ as well as height~$h_{c1}$ of the injection channel~$c1$. 
The resulting instance has been simulated with the proposed approach and, afterwards compared to the simulation results provided in~\cite{fink2020automatic}. The obtained results, i.e., the list of channels the payload droplet takes in the two examples, and subsequently its path, is visualized in Figure~\ref{fig:CaseStudyDroplets}. The simulation resulted in exactly the same droplet paths as reported in the CFD simulations by~\cite{fink2020automatic}. This confirms that the proposed modular and extendable simulation approach generates the same results with the same efficiency as dedicated and application-specific (but less flexible) solutions. 

% \begin{figure}[t]
%     \centering
%     \begin{subfigure}[b]{0.25\linewidth}
%         \centering
%         \includegraphics[width=\linewidth]{Images/CaseStudyC_1.png}
%         \caption{Experimental results from~\cite{chramiec_integrated_2020}}
%         \label{fig:CaseStudyMembrane1}
%     \end{subfigure}
%     \hspace{0.01\linewidth}
%     \begin{subfigure}[b]{0.25\linewidth}
%         \centering
%         \includegraphics[width=\linewidth]{Images/CaseStudyC_2.png}
%         \caption{CFD simulation results from~\cite{chramiec_integrated_2020}}
%         \label{fig:CaseStudyMembrane2}
%     \end{subfigure}
%     \begin{subfigure}[b]{0.25\linewidth}
%         \centering
%         \includegraphics[width=\linewidth]{Images/CaseStudyC_3.png}
%         \caption{Simulator results}
%         \label{fig:CaseStudyMembrane3}
%     \end{subfigure}
%     \vspace{0.2cm}
%     \caption{Concentration profiles in the OoC geometry, specifically in the connection channel and the two organ chambers. Subfigures (a) and (b) are reproduced from~\cite{chramiec_integrated_2020} with permission from the Royal Society of Chemistry}
%     \label{fig:CaseStudyMembrane}
% \end{figure}

\begin{figure} [t]
    \centering
    \includegraphics[width=0.85\linewidth]{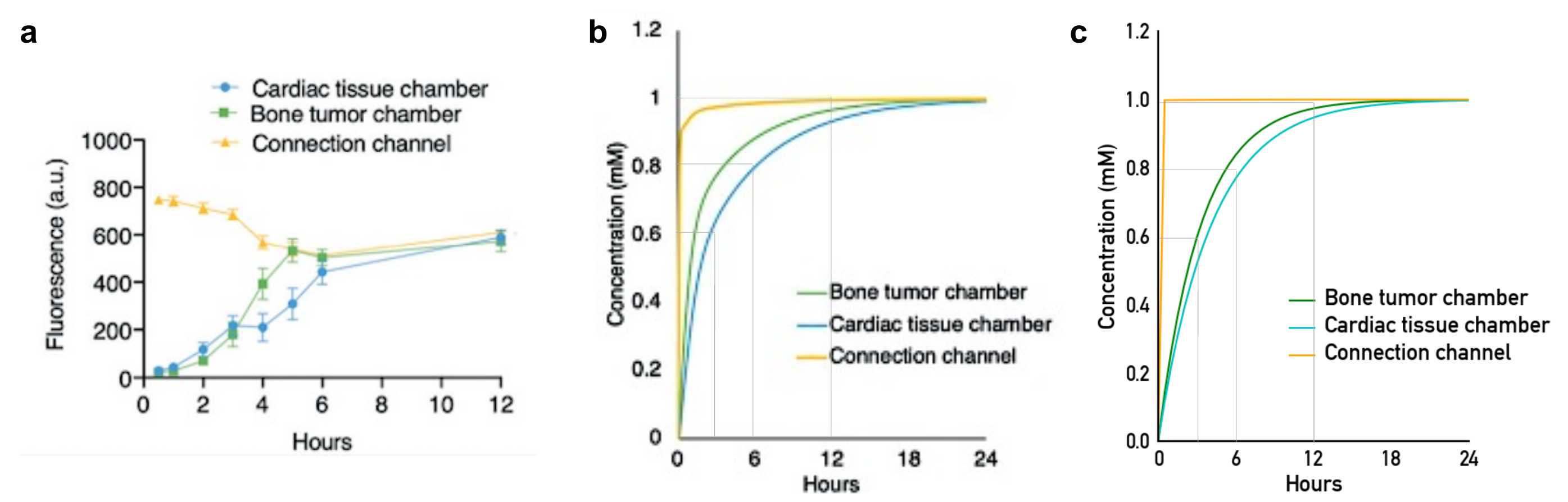}
    \caption{Concentration profiles in the OoC geometry, specifically in the connection channel and the two organ chambers. \textbf{a} Experimental results from~\cite{chramiec_integrated_2020}, \textbf{b} CFD simulation results from~\cite{chramiec_integrated_2020}, \textbf{c} Simulator results. \textbf{a}, and \textbf{b} are reproduced from~\cite{chramiec_integrated_2020} with permission from the Royal Society of Chemistry}
    \label{fig:CaseStudyMembrane}
\end{figure}

\subsection{Membranes}
As a final case study, we considered an \emph{Organ-on-Chip}~(OoC, \cite{chramiec_integrated_2020}) device which serves as suitable example to evaluate the membrane module extension. 
Membranes in microfluidic devices have become especially relevant for OoC devices. 
OoCs are testing platforms that contain miniaturized organ tissues to represent the physiology of an animal or human. The microfluidic network resembles the blood or fluid circulation in the body, and the organ tissues the organs. On-chip, they are often cultured in separate tanks that are connected via a membrane to the microfluidic network.

Abstract simulations of OoCs have so far been conducted with compartment models~\cite{jourdan_compartmental_2019} in which the geometry is separated into smaller parts that are individually defined and evaluated. However, they require a high set-up effort, which makes them less versatile.
Alternatively, when using the 1D simulation approach proposed in this work, both the flux of species or fluids across membranes as well as complex and easily adaptable microfluidic channel networks can be simulated. This is, to the best of our knowledge, the first time, this application has been considered for 1D simulation.

For the case study, the OoC set-up described in~\cite{chramiec_integrated_2020} has been considered, and the diffusion of a drug (linsitinib) through the membrane into the organ modules was simulated. Figure~\ref{fig:CaseStudyMembrane}b shows the originally provided data from CFD simulations that were experimentally verified using the fluorescent molecule FITC (Figure~\ref{fig:CaseStudyMembrane}a, the graph depicts the measured fluorescence values that are not yet normalized to concentration)~\cite{chramiec_integrated_2020}, as well as the data obtained by the simulator proposed in this work (Figure~\ref{fig:CaseStudyMembrane}c).
The model shows a very similar curve as well as the same uniform distribution of linsitinib after 12~hours. The MAE between the reported graph we extracted using graph analysis and the simulator results, at 0, 3, 6, 9, 12, 15, 18, 21, and 24~hours, is below 5~\% for the tumor and cardiac tissue and below 1~\% for the connection channel. Confirming that the proposed approach can generate the desired result with great accuracy. 

\bigskip 

Overall, the case studies illustrate the wide range of applicability of the proposed simulator and its extendability with various modules for the simulation of microfluidic applications. In fact, we were able to efficiently and correctly simulate the behavior of microflidic devices including mixing applications, droplets, and, for the first time, membranes using a single tool.
Moreover, the extendability allows to include more applications resulting in a comprehensive 1D simulator for microfluidics.
To this end, all implementations conducted in this work, along with test files, including additional network definitions, and a step-by-step guide, are also made publicly available as \mbox{open-source} at https://github.com/cda-tum/mmft-modular-1D-simulator, allowing users to easily adapt them for their own use cases.

\section{Conclusion}\label{sec:concl}
This work presented a modular and extendable 1D simulation for microfluidic devices. This way, complex microfluidic channel networks can be efficiently evaluated while saving computational resources and time due to the \mbox{high-level} abstraction. The tool allows for seamless integration of \mbox{application-specific} modules without the need to redevelop a new tool from scratch each time. Moreover, the proposed appraoch even allowed to support membranes in 1D simulation for the first time. Three representative \mbox{application-specific} aspects were implemented and compared with results from the literature. Case studies showed that this indeed allowed to efficiently simulate a broad spectrum of microfluidic applications that matches previous results or even fabricated devices and allowed to validate the simulations. 
While the current version of the software is fully functional and can be used with the provided step-by-step guide, further developments could explore the addition of a \emph{Graphical User Interface}~(GUI) to further enhance usability. 
The tool is available as an \mbox{open-source} software package at https://github.com/cda-tum/mmft-modular-1D-simulator. 

\section*{Data Availability Statement}
The code for the simulator tool is publicly available at https://github.com/cda-tum/mmft-modular-1D-simulator.

\bibliography{output.bbl}

\section*{Acknowledgements}
This work has partially been supported by the FFG project AUTOMATE (project number: 890068) as well as by BMK, BMDW, and the State of Upper Austria in the frame of the COMET Program managed by FFG.

\section*{Author Contributions}
M.E., F.C., and R.W. conceptualized the project. M.E. and F.C. developed the methodology and conducted the formal analysis. F.C., with support from M.E., worked on the software. M.E. wrote the original draft and was responsible for visualization. M.E. and R.W. supervised the project. R.W. reviewed and edited the manuscript.

\section*{Competing Interests}
The authors declare no competing interests.

\end{document}